# Multi-domain Reversible Data Hiding in JPEG Images

Zhaoxia Yin, Member, IEEE, Hongnian Guo, Yang Du

Abstract—As a branch of reversible data hiding (RDH), reversible data hiding in JEPG is particularly important. Because JPEG images are widely used, it is great significance to study reversible data hiding algorithm for JEPG images. The existing JEPG reversible data methods can be divided into two categories, one is based on Discrete Cosine Transform (DCT) coefficients modification, the other is based on Huffman table modification, the methods based on DCT coefficient modification result in large file expansion and visual quality distortion, while the methods based on entropy coding domain modification have low capacity and they may lead to large file expansion. In order to effectively solve the problems in these two kinds of methods, this paper proposes a reversible data hiding in JPEG images methods based on multi-domain modification. In this method, the secret data is divided into two parts by payload distribution algorithm, part of the secret data is first embedded in the DCT coefficient domain, and then the remaining secret data is embedded in the entropy coding domain. Experimental results demonstrate that most JPEG image files with this scheme have smaller file size increment and higher payload than previous RDH schemes.

# I. INTRODUCTION

Data hiding can embed the secret data bits into the carrier imperceptibly, and extract the secret information completely. Reversible data hiding (RDH) [1] can not only extract the secret information completely, but also recover the original carrier losslessly. Therefore, reversible data hiding technology can be applied to medical, military and other fields which have high requirements for the lossless recovery of the original carrier.

The research for reversible data hiding methods is not only for spatial image, but also for JPEG image. There are three kinds of reversible data hiding methods for spatial images: lossless compression [2] [3], difference expansion (DE) [4] [5] and histogram shifting (HS) [6]-[8]. Because JPEG image is the most popular image format, it is necessary to study the reversible data hiding in JPEG images.

So far, two category methods applied to reversible data hiding in JPEG images. The first category methods is based on modifying the Discrete Cosine Transform (DCT) coefficient domain. The second category methods is based on modifying entropy coding domain. The method based on DCT coefficient modification is first proposed by Friedrich [9], who uses lossless compression algorithm to compress the least significant bit (LSB) of AC coefficient, so that the redundant space is used to embed secret information. In addition, Friedrich proposes that the scaling factor is set to 2, that is, the quantization step is divided by 2, and the corresponding quantization AC coefficient is multiplied by 2 so that secret data can be embedded. Wang et al. [10] improve Friedrich's method, sets the scaling factor to k, and determines the modification range of quantization table according to the unit distortion at different positions in the 8*8 block. Xuan et al. [11] use histogram shifting strategy to embed secret data. Huang et al. [12] embed the secret data into ±1 non-zero AC coefficients, and uses the block selection strategy to improve the embedding capacity and reduce the file expansion. Wedaj et al. [13] propose embedding efficiency to measure each frequency. Hou et al. [14] propose the embedding cost to define the embedding distortion of each frequency. The frequency with small embedding distortion is selected to embed secret data, which effectively increases the peak signal-to-noise ratio (PSNR). He et al. [15] propose negative influence models, optimizing PSNR and file expansion at the same time, the file expansion is effectively reduced and the peak signal-to-noise ratio is improved. He et al. [16] improve block sort and embed data in blocks that result in less distortion. Yin et al.[17] propose multi-objective optimization, which measures distortion and expansion, and optimizes visual quality and file expansion at the same time. Xiao et al. [18] modify multiple histogram, and propose effective block smoothing and frequency smoothing to improve performance. Those method based on DCT coefficient modification are also called reversible data hiding. The method based on Huffman table is first proposed by Mobasseri et al. [19] And then improved by Qian and Zhang et al. [20] This kind of method embeds secret data by establishing the mapping between the used and unused VLCs codes of the same length. Qiu et al. [21] improve the embedding capacity by VLC frequency ordering and mapping optimizing. After embedding the secret data, the visual quality and file size of the carrier image remain unchanged. One drawback of these methods is the low embedding capacity. Du et al. [22] propose a new method based on Huffman table modification in 2018. This method improves the embedding capacity while keeping the visual quality unchanged, but there is a certain amount of file expansion. In 2020, they [23] further construct a new mapping method to reduce file expansion through the proposed general VLC mapping (GVM). Those method based on Huffman table modification are also called lossless data hiding.

Reversible data hiding in JPEG images method based on DCT coefficient domain modification, it causes the image visual quality distortion and file expansion, but it creates larger payload. Reversible data hiding in JPEG images method based on entropy coding domain modification, it does not cause image visual quality distortion, but the payload is limited and it may cause file expansion. Based on the above, this paper proposes a new reversible data hiding method. This method takes advantages of multi-domain, it results in smaller file expansion and higher PSNR than the methods which based on DCT coefficient domain modification, and it has smaller file expansion compared with the methods based on entropy coding domain modification.

The remainder of this paper is organized as follows. In second II, the preliminaries work is introduced, including the knowledge of DCT coefficient domain and entropy coding domain. The process of this method is presented in the III section, including the embedding method used in the two domains and the payload distribution algorithm. The experimental results and analysis are proposed in Section IV. Finally, this paper is concluded in Section V.

# II. PRELIMINARIES

In this section, we first introduce the JPEG compression process, and then introduce the reversible data hiding in JPEG images method in DCT coefficient domain and entropy coding domain.

A. JPEG compression process

JPEG compression process is shown in the figure 1. An original image is divided into several 8x8 blocks, and DCT is performed for each block, then each block is quantified with the quantization table, and the quantized DCT coefficients are encoded into an intermediate format, in which the DC coefficients are encoded by differential pulse code modulation (DPCM) and the AC coefficients are encoded by run-length encoding (RLE). Take the AC coefficient as an example, a AC coefficient is encoded as run/length, value (RLV), where variable length integer encoding (VLI) is used to encode value and run/length is encoded by Huffman coding to get variable-length code (VLC). The compressed JPEG bitstream is obtained by combining VLC codes with VLI codes.

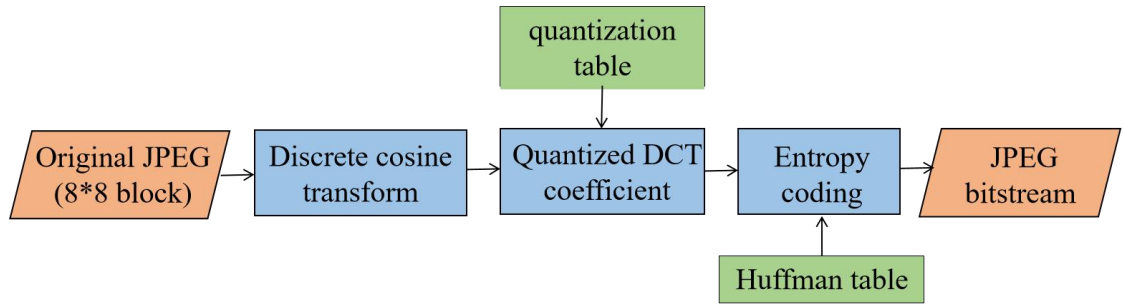

Fig.1 JPEG compression process

Since the proposed method need modify Huffman table, we will introduce Huffman table saved in JPEG header file. Many segments are contained in JPEG header file, in which DHT segment contains Huffman table information. Each VLC corresponds to the RLV of AC coefficient in entropy coding. VLCs is encoded by Huffman Coding. Huffman coding rule is that the shorter code is assigned to the RLV with higher frequency, and the longer code is assigned to the RLV with lower frequency. For example, if the *run/length* is "0/4", the corresponding VLC is "1011"; if the *run/length* is "0/2", the corresponding VLC is "01". AC coefficients correspond to 162 VLCs, including "0/0", "F/0" and 160 VLCs corresponding to *run/length* from "0/1" to "F/A". The length of the VLCs is between 2 and 16 bits. Here F and A are hexadecimal.

B. Modification method in two domain

a. Modification method in DCT coefficient domain

First of all, we introduce the method based on DCT coefficient domain modification here. Huang [12] et al. propose to embed the secret data into ±1 non-zero AC coefficients, He [15] et al. and Yin [17] et al. put forward some improvements on the basis of it, He et al's method is one of the most advanced methods in the method based on DCT coefficient domain modification. The embedding formula of Huang et al's method is as follows.

$$\overline{d} = \begin{cases} d + sign(d)*b & if\ |d|=1 \\ d + sign(d) & if\ |d|>1 \end{cases}$$

$$where\quad sign(x) = \begin{cases} 1 & if\ x>0 \\ 0 & if\ x=0 \\ -1 & if\ x<0 \end{cases}$$

(1)

Where $d$ is the quantized DCT coefficient, $\overline{d}$ is the modified DCT coefficient, b is the secret data bit to be embedded. And the extracting and recovering formula of Huang et al's method is as follows.

$$b = \begin{cases} 0, & if\ |\overline{d}|=1 \\ 1, & if\ |\overline{d}|=2 \end{cases}$$

(2)

$$d = \begin{cases} sign(\overline{d}), & if\ 1 \leq |\overline{d}| \leq 2 \\ \overline{d} - sign(\overline{d}), & if\ |\overline{d}| \geq 3 \end{cases}$$

(3)

b. Modification method in entropy coding domain

Next, we will introduce the method based on entropy coding domain modification. Most of the previous methods based on entropy coding domain modification have no visual quality distortion and no file expansion, but the payload is limited.

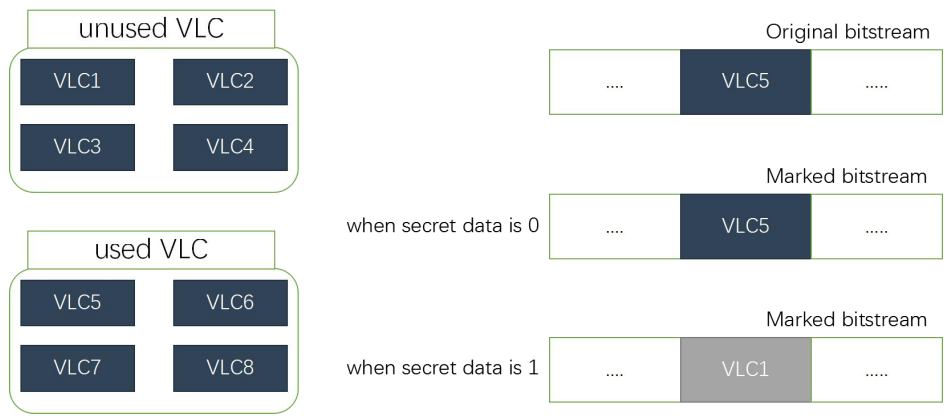

Fig.2 Modification method in entropy coding domain

In 2009, Mobasseri et al. [19] first proposed the concept of VLC mapping, Qian and Zhang et al. [20] propose the method that a one to many mapping relationship between unused VLCs and used VLCs, which not only keeps the visual quality and file expansion unchanged, but also greatly increases payload, But the payload is still low compared with the method based on DCT coefficient domain modification. Therefore, Du et al. [22] propose a new method based on entropy coding domain modification, its essence is to use unused VLCs instead of used VLCs to embed secret data. Du et al. [22] are not limited to VLCs with the same code length, so there will be file expansion, and use HS strategy. The principle of modification in entropy coding domain is shown in figure 2. VLCs in bitstream are divided into two categories: used VLCs and unused VLCs. If the secret data to be embedded is 0, the VLC remains unchanged, if the secret data to be embedded is 1, the used VLCs is replaced by the unused VLCs.

## III. PROPOSED SCHEME

In this section, the proposed method is described in detail. The framework of the proposed scheme is illustrated in figure 3. The secret data is divided into two parts by payload distribution algorithm, part of the secret data is first embedded in the DCT coefficient domain, and then the remaining secret data is embedded in the entropy coding domain. How to use the advantages of the two domains to distribute secret data is the key to affect the experimental results.

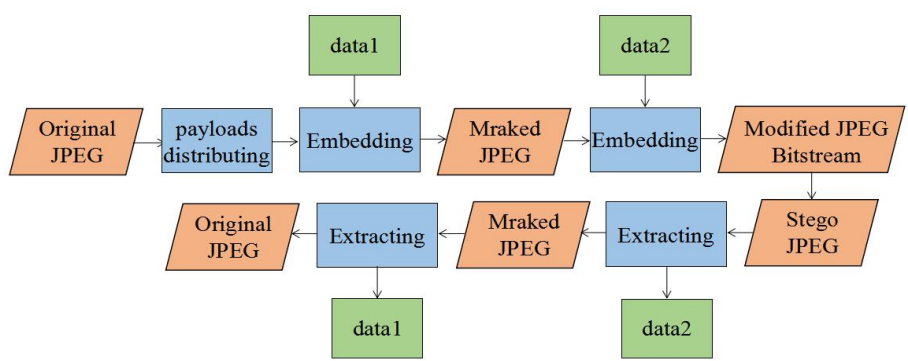

Fig.3 The framework of the proposed scheme

A. Data Embedding & Extraction

Because the secret data distributed to two domains needs to be embedded with two ways, two embedding methods will be proposed here. Part of secret data is first embedded in the DCT coefficient domain, and the marked image is obtained, remaining secret data is embedded with method base on entropy coding domain modification in the marked image. How to distribute payload is described in Section III-B.

a. Embedding method in DCT coefficient domain

Among the methods based on DCT coefficient domain modification, He et al. [15] achieve good results through the negative influence model. Inspired by He et al's [15] method, we use the method similar to He et al's [15] method, that is, the secret data is embedded into the non-zero AC coefficients where frequency is selected by frequency selection strategy. Frequency selection strategy will be introduced in Section III-B. The detailed steps of embedding secret data in JPEG image are described below. Secret data bit length is *data*.

1) Decode the original grayscale JPEG image (denote by $I$) to obtain the quantized DCT coefficient blocks $D$.
2) Sort the DCT blocks $D$ in descending order according to the number of zero-valued AC coefficients.
3) The first $k$ frequencies are selected according to the distributed secret data bit length (denote by $L1$) and frequency selection strategy. Select as few blocks as possible from the sorted DCT blocks obtained in 2) according to the distributed secret data length and $k$.
4) Embed secret data in non-zero AC coefficients whose frequency belongs to the set of $k$ according to formula (1).
5) Let $k=k+1$, repeat the Step 4) and Step 5) until $k=63$. The minimum file expansion of k and the corresponding number of used blocks are denoted $\bar{k}$ and $\bar{n}$.
6) Let *run* = 0, embed secret data in each of the non-zero AC coefficients whose frequency belongs to the set of $\bar{k}$ and zero-run length equals to run in the chosen $\bar{n}$ blocks. Then set *run=run+1*, repeat the Step 6) until all secret data bits are embedded. Modified quantized DCT coefficients are encoded to obtain the marked image. The last zero-run used during data embedding is represented by $\overline{run}$.

b. Embedding method in entropy coding domain

Most of the methods based on entropy coding domain modification have no file expansion, but the payload is very limited. A novel method based on entropy coding domain modification proposed by Du [22] et al. in 2018. This method improves the payload greatly. Even if it leads to a certain file expansion, it is acceptable. [22] obtains VLC frequency distribution histogram by analyzing entropy coding stream, and then shifts all VLCs to the right except the first VLC. The first VLC is replaced by the second VLC to embed secret data.

In this paper, we improve the method of [22]. The Lena image which quality factor is 50 is taken as an example. We sort the VLCs in VLC frequency distribution histogram according to the VLC frequency, this process is shown in the figure 4, and then shift the histogram according to the length of secret data, finally embed secret data, this process is shown in the figure 5. The detailed steps are as follows.

1) Parse the DHT segment and entropy coding bitstream of JPEG image, count the number of *VLC* in entropy coding bitstream, and then generate VLC frequency distribution histogram according to the sequence of *run/length* in DHT segment.

2) Sort the VLCs in VLC frequency distribution histogram according to the *VLC* frequency, and modify DHT segment. It is shown in figure 4.

3) From right to left, the first *VLC* whose frequency is larger than the distributed secret data bit length (denote by *L2*) is selected as the peak point *P*, and then from *P* to the right, the first *VLC* whose frequency is 0 is selected as the zero point *Z*, then shift all *VLC* from *P + 1 to Z*. Similarly, modify DHT segment. *Z* is adjacent to *P* and *Z* is to the right of *P* after shifting, it can be easily seen in figure 5(a).

4) Scan entropy coding bitstream, when the *VLC* whose position is at *P* is scanned, then this *VLC* is used to embed secret data. If the secret data to be embedded is "0", then the *VLC* remains unchanged. If the secret data to be embedded is "1", then the *VLC* is modified to the *VLC* corresponding position is at *P* + 1. it can be seen in figure 5(b). When all the secret data bit are embedded, scanning entropy coding bitstream is stopped and JPEG bitstream is generated.

c. Data extraction and image recovery

When all secret data bit are embedded, the marked JPEG bitstream is generated. We first extract the secret data embedded in the entropy coding domain, and then extract the secret data embedded in the DCT coefficient domain. The process of secret data extraction and original image recovery is as follows.

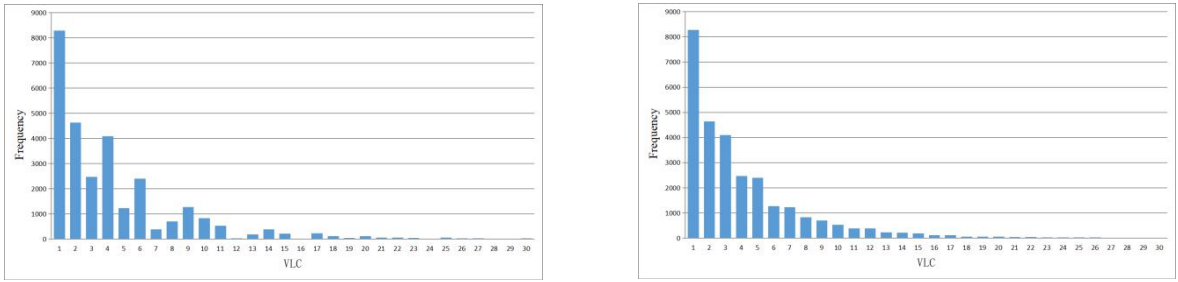

(a): before sorting  (b): after sorting

Fig.4 The process of sorting the VLCs ((a): before sorting; (b): after sorting)

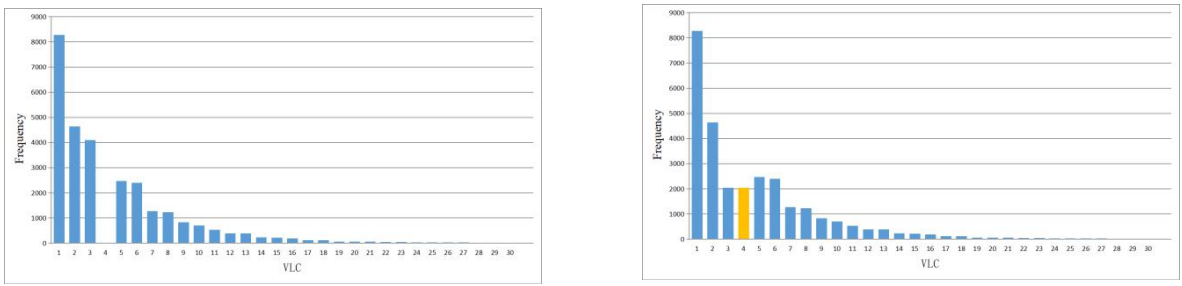

(a): shifting histogram  (b): embedding secret data

Fig.5 The process of modifying histogram ((a): shifting histogram ; (b): embedding secret data)

1) Parse the DHT segment and entropy coding segment of JPEG image, calculate the frequency of VLC in the coding segment, and then generate the VLC frequency distribution histogram according to the sequence of *run/length* in DHT segment.

2) Find two VLCs with the same *run/length*, i.e. peak point $P$ and zero point $Z$.

3) Scan the entropy coding segment, extract secret data according to $L2$, $P$ and $Z$ until all secret data bits are extracted, and stop scanning.

4) According to the standard Huffman table, the *run/length* sequence is modified to the original *run/length* sequence, and the corresponding VLC is modified to generate the recovered JPEG marked image.

5) Decode the marked JPEG image to obtain DCT coefficients matrix, the DCT coefficients are divided into 8*8 blocks $D'$, and they are sorted in ascending order according to the number of zero-valued AC coefficients.

6) As few blocks as possible are selected from the sorted DCT coefficient blocks obtained in Step 5) according to the $\overline{run}$ and $L2$.

7) Set *run* = 0, extract the embedded data using (2) from each of the nonzero AC coefficients whose frequency belongs to the set of $\bar{k}$ selected frequencies and zero-run length equals to *run* in unused DCT block. And each non-zero AC coefficient is recovered by formula (3).

8) Set *run=run+1* and repeat step 7) until all the data bits are extracted. The recovered image can be generated.

B. Payload distribution

Because there are many reversible data hiding technologies for JPEG images, the embedding methods for DCT coefficient domain and entropy coding domain are very different, and the results is also different. For reversible data hiding in JEPG based on multi-domain modification, payload distribution algorithm is the key to affect the results. We introduce the method in two domains above, the payload distribution algorithm is described below.

a. Embedding cost

Inspired by He et al. [15] we propose the embedding cost for two domains, the payload is distributed to two domains according to embedding cost.

It is worth noting that the visual quality is not taken into account in the calculation of embedding cost in the proposed method, there are two reasons: one is that the modification in entropy coding domain does not cause image visual distortion, the calculation of visual distortion is of little significance; on the other hand, we distribute all secret data bits to two domains, and the modification in entropy coding domain does not cause vision distortion, only the modification in DCT coefficient domain cause visual distortion, so the visual quality must be better than the previous methods based on DCT coefficient domain modification.

When DCT coefficients are modified, the length of VLC codes corresponding to AC coefficients will change, which is the main reason for file expansion. Huffman code length table can be generated according to DHT segment in JPEG header file, and Huffman code length increments table can be calculated according to the Huffman code length table, Huffman code length table is shown in the figure 6, Huffman code length increments table is shown in the figure 7. Embedding cost in DCT coefficient domain can be calculated as follow.

$$S(k)=\sum_{j=1}^{N}\omega S_j(k) \qquad (5)$$

$$S_j(k)=\begin{cases} hcit[r_j(k),c_j(k)]+1, & if\ |d(k)|=2^z-1 \\ 0, & other \end{cases} \qquad (6)$$

$$\omega=\begin{cases} 0, & if\ d_j(k)=0 \\ \frac{1}{2}, & if\ |d_j(k)|=1 \\ 1, & if\ |d_j(k)|>1 \end{cases} \qquad (7)$$

| R/L | 1 | 2 | 3 | 4 | 5 | 6 | 7 | 8 | 9 | 10 |
|---|---|---|---|---|---|---|---|---|---|---|
| 0 | 2 | 2 | 3 | 4 | 5 | 7 | 8 | 10 | 16 | 16 |
| 1 | 4 | 5 | 7 | 9 | 11 | 16 | 16 | 16 | 16 | 16 |
| 2 | 5 | 8 | 10 | 12 | 16 | 16 | 16 | 16 | 16 | 16 |
| 3 | 6 | 9 | 12 | 16 | 16 | 16 | 16 | 16 | 16 | 16 |
| 4 | 6 | 10 | 16 | 16 | 16 | 16 | 16 | 16 | 16 | 16 |
| 5 | 7 | 11 | 16 | 16 | 16 | 16 | 16 | 16 | 16 | 16 |
| 6 | 7 | 12 | 16 | 16 | 16 | 16 | 16 | 16 | 16 | 16 |
| 7 | 8 | 12 | 16 | 16 | 16 | 16 | 16 | 16 | 16 | 16 |
| 8 | 9 | 15 | 16 | 16 | 16 | 16 | 16 | 16 | 16 | 16 |
| 9 | 9 | 16 | 16 | 16 | 16 | 16 | 16 | 16 | 16 | 16 |
| 10 | 9 | 16 | 16 | 16 | 16 | 16 | 16 | 16 | 16 | 16 |
| 11 | 9 | 16 | 16 | 16 | 16 | 16 | 16 | 16 | 16 | 16 |
| 12 | 10 | 16 | 16 | 16 | 16 | 16 | 16 | 16 | 16 | 16 |
| 13 | 11 | 16 | 16 | 16 | 16 | 16 | 16 | 16 | 16 | 16 |
| 14 | 16 | 16 | 16 | 16 | 16 | 16 | 16 | 16 | 16 | 16 |
| 15 | 16 | 16 | 16 | 16 | 16 | 16 | 16 | 16 | 16 | 16 |

Fig.6 Huffman code length table

| R/L | 1 | 2 | 3 | 4 | 5 | 6 | 7 | 8 | 9 |
|---|---|---|---|---|---|---|---|---|---|
| 0 | 0 | 1 | 1 | 1 | 2 | 1 | 2 | 6 | 0 |
| 1 | 1 | 2 | 2 | 2 | 5 | 0 | 0 | 0 | 0 |
| 2 | 3 | 2 | 2 | 4 | 0 | 0 | 0 | 0 | 0 |
| 3 | 3 | 3 | 4 | 0 | 0 | 0 | 0 | 0 | 0 |
| 4 | 4 | 6 | 0 | 0 | 0 | 0 | 0 | 0 | 0 |
| 5 | 4 | 5 | 0 | 0 | 0 | 0 | 0 | 0 | 0 |
| 6 | 5 | 4 | 0 | 0 | 0 | 0 | 0 | 0 | 0 |
| 7 | 4 | 4 | 0 | 0 | 0 | 0 | 0 | 0 | 0 |
| 8 | 6 | 1 | 0 | 0 | 0 | 0 | 0 | 0 | 0 |
| 9 | 7 | 16 | 0 | 0 | 0 | 0 | 0 | 0 | 0 |
| 10 | 7 | 16 | 0 | 0 | 0 | 0 | 0 | 0 | 0 |
| 11 | 6 | 16 | 0 | 0 | 0 | 0 | 0 | 0 | 0 |
| 12 | 6 | 16 | 0 | 0 | 0 | 0 | 0 | 0 | 0 |
| 13 | 5 | 16 | 0 | 0 | 0 | 0 | 0 | 0 | 0 |
| 14 | 0 | 16 | 0 | 0 | 0 | 0 | 0 | 0 | 0 |
| 15 | 0 | 16 | 0 | 0 | 0 | 0 | 0 | 0 | 0 |

Fig.7 Huffman code length increments table

Where *hcit* is the Huffman code length increments table, r and c denote the run and length of the AC coefficient, $N$ is the number of 8*8 blocks in the image. $S(k)$ is the total file expansion caused by modifying non-zero AC coefficient which frequency is the $k$ in all N blocks, $k$ ($2\leq k\leq 64$) is the zigzag scanning order of the 8 * 8 matrix.

$$UF(k)=\frac{S(k)}{L(k)} \qquad (8)$$

Where $L(k)$ is the number of nonzero AC coefficients which frequency is the $k$, $UF(k)$ is embedding cost. It can be proved from (8) that the file expansion when needed frequency is $k$, denote by *INC1*, can be calculated as follow.

$$INC1(k)=UF(k)\times L(k) \qquad (9)$$

As can be seen from Section III-A, after VLCs are sorted according to the VLC frequency, the proposed embedding method in entropy coding domain can cause file expansion, file expansion is caused by two operations, one is the shifting VLC frequency distribution histogram, the other is the embedding secret data, it can be seen from figure 4 too. Embedding cost in entropy coding domain can be calculated as follow,

$$S(P) = \sum_{i=P+1}^{Z} [sum(\overline{VLC_i}) - sum(VLC_i)] L_{VLC} \qquad (10)$$

where $S(P)$ is the file expansion caused by histogram shifting when $P$ is selected as the peak point, $Z$ is zero point, $VLC_i$ is the VLC whose position is at $i$, $\overline{VLC_i}$ is the VLC whose position is at i after histogram shifting. $sum$ is the number of $VLC_i$, $L_{VLC_i}$ is the length of the $VLC_i$,

$$M(P) = \frac{1}{2} sum(VLC_P)[L(VLC_{P+1}) - L(VLC_P)] \qquad (11)$$

where $M(P)$ is the file expansion caused by embedding secret data when $P$ is selected as the peak point. The default length of secret data bits is the peak height here. Embedding cost in entropy coding domain, denote by $E(P)$, can be calculated by

$$E(P) = \frac{S(P) + M(P)}{sum(VLC_P)} \qquad (12)$$

It can be proved from (11) that the file expansion at peak point $P$, denote by $INC2$, can be calculated as follow.

$$INC2(P) = E(P) \times sum(VLC_p) \qquad (13)$$

$E$ and $INC2$ are function of peak point $P$, $UF$ and $INC1$ are function of frequency $k$, we choose the best frequency and peak point according to $UF$ and $E$, in other words, we distribute payload according to $UF$ and $E$.

b. Payload distribution and judgment condition

We get $UF$ and $E$ above. In order to minimize the file expansion result, We distribute the payload according to $UF$ and $E$. We calculate the embedding cost in entropy coding domain by (11), the default $L2$ is the peak height, that is $sum(VLC_P)$, and in experiments, we find that the file expansion caused by histogram shifting is more than that caused by embedding secret data. Therefore, $L2$ needs to be as close as possible to $sum(VLC_P)$ so that file expansion is smaller. $UF$ and $E$ are sorted in ascending order.

First, the initial values of $L1$ and $L2$ can be calculated by algorithm 1, we also get $P$ and $k$ from algorithm 1, in this case, $E(P)$ is less than $UF(k)$. Now there is a question around us when we get $L1$ and *L2* from algorithm 1. After the *L1* secret data bits are embedded in DCT coefficient domain, the embedding cost in entropy coding domain would be changed. As we know, *run/length* is encoded using Huffman code. For example, a non-zero AC coefficient value is 1 whose *run* is 0 and *run/length* is (0/1). Its VLC code is 00, it would be modified to 01 if secret data bit to be embedded is "1". When the VLCs are modified, embedding cost would be changed. Therefore, we need to know whether the selected peak point *P* is reasonable or not.

To solve this problem, we put forward a judgment condition. This condition is to judge whether the peak point is reasonable or not. When the selected peak point *P* meet the judgment condition, we say that the peak point is reasonable. When the peak point passes the detection of judgment condition, the peak point can be determined.

**Algorithm** 1: initializing payload distribution
**Input**: *UF, E, length*
**Output**: *L1, L2, P and k*
1. **begin**
2.    **for** *i*=2 to 64 **do**
3.       **for** *j*=1 to 162 **do**
4.          **if**   *E(j)>UF (i)*
5.             break;
6.          **end**
7.       **end**
8.       **if**   *UF (i)+sum(VLC$_j$)>length*
9.          break;
10.       **end**
11.    **end**
12.   *P=j*;
13.   *k=i*;
14.   *L2=sum(VLC$_P$)*;
15.   *L1=length-L2*;
16. **end**

After initializing load distribution payload, *L1* secret data bits are embedded in DCT coefficient domain firstly by step 1) to step 6) in Section III-B-a, we generate a marked image *I'*, as shown in figure 3. Because the embedding cost has changed, we get the embedding cost of marked image by (10), (11) and (12), denote by *E'*, *E'* is sorted in ascending order. After optimizing payload distribution, because the position of the peak is changed, the peak point *P* in *E'* is *j'*. The peak point with the largest frequency among all peak points whose embedding costs are less than that of *j'* is *j"*. The judgment condition is that whether the file expansion caused by embedding *L2-sum(VLC$_{j"}$)* secret data bits in entropy coding domain higher than that caused by embedding $L2$-$sum(VLC_{j''})$ secret data bits in DCT coefficient domain. If $L2$-$sum(VLC_{j''})$ secret data bits are embedded in entropy coding domain, we can calculate the resulting file expansion, denote by $S1$, according to (13) as follow.

$$S1 = E'(j') \times data2 - E'(j'') \times sum(VLC_{j''}) \tag{14}$$

Where $VLC_{j''}$ is the $VLC$ in marked image $I'$. And if $L2$-$sum(VLC_{j''})$ secret data bits are embedded in DCT coefficient domain, and the number of selected frequency is $k'$. we can calculate the resulting file expansion, denote by S2, according to (9) as follow.

$$S2 = INC1(k') - INC1(k) \tag{15}$$

If $S1 > S2$, we say that the selected peak point $P$ is unreasonable and select $j''$ as the peak point. If $S1 < S2$, the payload distribution is reasonable.

The peak point has been determined after distribute payload and judge condition in Section III-B-c, we need to optimize the payload distribution. Because the VLCs of entropy coding domain will change after the secret data bits are embedded in DCT coefficient domain. As we have analyzed above, We should optimize the payload distribution so that $L2$ is under and close to the height of selected peak point. We can embed $L1$ secret data bits in DCT coefficient domain, then compare with $sum(VLC_{j'})$ and $L2$, and then adjust the values of $L1$ and $L2$ so that $L2$ is under and close to the height of selected peak point. It is worth noting that before judging whether the selected peak point is reasonable, we should optimize payload distribution so that judgment condition is more accurately.

## IV. EXPERIMENTAL RESULTS

In this section, we show the results. The experimental results are divided into three parts, one is the test results of a single image, one is the test results of database(UCID) [24], and the last is the verification experiment. In order to demonstrate the advancement of the proposed method, the methods in [15], [17], [18] and [23] are compared. Method in [15], [17], [18] are more advanced in reversible data hiding methods. Du et al's [23] method is one of the most advanced method in lossless data hiding methods. We don't compare with Du et al's [22] method which had done in 2018, because his latest method [23] is significantly better than his previous method [22]. We use the JPEG image which quality factor is 50, 70 and 90, and Huffman table is default. The indexes we compare are file expansion, peak signal to noise ratio (PSNR). Since Du et al's [23] method is lossless data hiding, we do not compare PSNR with it.

A. Single image test

In the experiment, we first select two common grayscale images to test, including Baboon, Lena. These test images are shown in figure 7. The experimental results are shown in the Table I, Table II, Table III, Table IV, Table V and Table VI. We make bold the font of the data that has the advantage. It is worth mentioning that our method uses the file size change model proposed by He et al. in [15]. To show the effectiveness of our method, the parameter α in He et al's [15] method is set to 1.

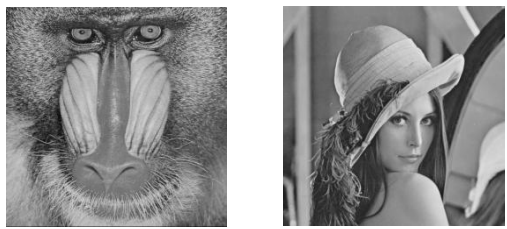

Fig.8 The test images

It can be seen from Table I to Table III that when the payload is the same, the file expansion of proposed method is smaller than all methods [15], [17], [18], [23] in most cases obviously, It can be seen that the performance of Du et al's [23] latest method is quite advanced. This shows the effectiveness of proposed embedded cost. The results in the Table I, Table II and Table III show that the file expansion of proposed and Du et al's [23] method is negative, the reason of this phenomenon is that the proposed method and Du et al's [23] method has the process of optimizing Huffman table. With the increase of quality factor, the file expansion advantage of the proposed method is no longer as obvious as that of low quality factor. It can be seen from Table IV to Table VI that the PSNR of proposed method is larger than He et al's method. The PSNR of Xiao et al's [18] and Yin et al's [17] method is better than that of the proposed method in some cases. However, the PSNR of the proposed method is better than that of all other methods in some cases, it can be see from Table IV. With the increase of quality factor, the PSNR of the proposed method is superior only at high payload, it can be seen from Table V and Table VI. In test result, the PSNR and file expansion of the proposed method are better than those of He et al's [15] method. This result shows that the proposed method is superior to that of He et al's [17] method, this shows that the proposed multi-domain is effective.

TABLE I: The file size changes in bits of the marked images with quality factor is 50 and embedded with different lengths of data using our method and four state-of-the-art methods.

| Image | Scheme | Payload | | | | | | | | |
|---|---|---|---|---|---|---|---|---|---|---|
| | | 2000 | 3500 | 5000 | 6500 | 8000 | 9500 | 11000 | 12500 | 14000 |
| Baboon | He et al. | 1568 | 3048 | 4360 | 6400 | 8232 | 10008 | 12136 | 14192 | 16096 |
| | Du et al. | -984 | 480 | 888 | 4424 | 7264 | 7584 | 12672 | 13144 | 13760 |
| | Xiao et al. | 2504 | 4128 | 5336 | 7952 | 9856 | 11688 | 14008 | 15856 | 18384 |
| | Yin et al. | 2656 | 4680 | 6544 | 8688 | 10768 | 12672 | 14704 | 17128 | 19432 |
| | Ours | **-3512** | **-2184** | **840** | **1304** | **2568** | **4440** | **5832** | **8056** | **9416** |
| Lena | He et al. | 1672 | 3464 | 5416 | 7480 | 9576 | 11888 | 14640 | 17256 | 20288 |
| | Du et al. | 512 | 1832 | 5008 | 6728 | 8800 | 10488 | 12632 | 14704 | 16752 |
| | Xiao et al. | 2728 | 4752 | 6624 | 8160 | 10152 | 12008 | 14792 | 17664 | 20344 |
| | Yin et al. | 2896 | 4928 | 6864 | 8616 | 11048 | 13048 | 15520 | 17928 | 20936 |
| | Ours | **-840** | **904** | **2616** | **4464** | **6384** | **8328** | **10216** | **14864** | **14232** |

TABLE II: The file size changes in bits of the marked images with quality factor is 70 and embedded with different lengths of data using our method and four state-of-the-art methods.

| Image | Scheme | Payload | | | | | | | | |
|---|---|---|---|---|---|---|---|---|---|---|
| | | 2000 | 4000 | 6000 | 8000 | 10000 | 12000 | 14000 | 16000 | 18000 |
| Baboon | He et al. | 2176 | 3936 | 5712 | 8520 | 11144 | 13856 | 16584 | 19960 | 22416 |
| | Du et al. | 1024 | 5248 | 6192 | 10656 | 11624 | 12664 | 21848 | 22056 | 25088 |
| | Xiao et al. | 2408 | 4664 | 7784 | 10192 | 13128 | 15992 | 19352 | 23136 | 26240 |
| | Yin et al. | 3160 | 5568 | 8664 | 11376 | 14144 | 17208 | 20800 | 24464 | 27536 |
| | Ours | **-656** | **736** | **2928** | **5408** | **9000** | **10760** | **13680** | **16336** | **18016** |
| Lena | He et al. | 1592 | 3816 | 6456 | 9872 | 12272 | 15336 | 18608 | 22272 | 26024 |
| | Du et al. | 1976 | 5216 | 7680 | 9128 | 13520 | 16256 | 18744 | 23512 | 26440 |
| | Xiao et al. | 2824 | 5448 | 7792 | 10704 | 12976 | 15712 | 18520 | 22416 | 26024 |
| | Yin et al. | 2896 | 5808 | 8592 | 11808 | 14208 | 16984 | 19720 | 23184 | 26912 |
| | Ours | **728** | **3400** | **5136** | **7792** | **10144** | **12360** | **14992** | **20376** | **23264** |

TABLE III: The file size changes in bits of the marked images with quality factor is 90 and embedded with different lengths of data using our method and four state-of-the-art methods.

| Image | Scheme | Payload | | | | | | | | |
|---|---|---|---|---|---|---|---|---|---|---|
| | | 2000 | 5000 | 8000 | 11000 | 14000 | 17000 | 20000 | 23000 | 26000 |
| Baboon | He et al. | 1904 | 4928 | 8136 | 11648 | 14984 | 19680 | 22984 | 27192 | 30512 |
| | Du et al. | -5992 | -2128 | 3768 | 7592 | 8144 | 19352 | 18704 | 23728 | 24400 |
| | Xiao et al. | 2400 | 6512 | 10488 | 15872 | 20480 | 24800 | 28184 | 33640 | 40056 |
| | Yin et al. | 2976 | 8056 | 13224 | 18536 | 23984 | 28224 | 34176 | 39040 | 44416 |
| | Ours | **-7744** | **-4632** | **-1544** | **1984** | **5480** | **8352** | **11888** | **15536** | **19256** |
| Lena | He et al. | 1808 | 4984 | 8368 | 12224 | 16112 | 20448 | 24464 | **28896** | 33472 |
| | Du et al. | 2456 | 8752 | 13968 | 18952 | 24472 | 29112 | 30152 | 41368 | 43264 |
| | Xiao et al. | 2328 | 5536 | 9064 | 11720 | 16592 | 20344 | 24944 | 29824 | 35184 |
| | Yin et al. | 2960 | 6296 | 10680 | 14776 | 19464 | 23600 | 28472 | 33008 | 38480 |
| | Ours | **112** | **3408** | **7352** | **10272** | **14800** | **18656** | **23792** | 29704 | **30328** |

TABLE IV: The PSNR values of the marked images with quality factor is 50 and embedded with different lengths of data using our method and three state-of-the-art methods.

| Image | Scheme | Payload | | | | | | | | |
|---|---|---|---|---|---|---|---|---|---|---|
| | | 2000 | 3500 | 5000 | 6500 | 8000 | 9500 | 11000 | 12500 | 14000 |
| Baboon | He et al. | 39.99 | 38.77 | 42.98 | 35.83 | 34.55 | 34.03 | 33.35 | 32.80 | 32.38 |
| | Xiao et al. | **48.00** | **45.01** | 36.66 | 41.23 | 39.87 | **38.76** | **37.82** | **36.84** | **36.00** |
| | Yin et al. | 47.80 | 44.83 | 42.88 | 41.15 | 39.83 | 38.63 | 37.55 | 36.62 | 35.67 |
| | Ours | 42.31 | 38.79 | **77.97** | **43.89** | **42.04** | 38.29 | 36.00 | 35.20 | 34.02 |
| Lena | He et al. | 45.33 | 43.68 | 41.04 | 39.70 | 39.10 | 38.09 | 37.41 | 36.81 | 36.06 |
| | Xiao et al. | **49.53** | **46.38** | 44.23 | 42.63 | 41.16 | 39.82 | 38.71 | 37.39 | 36.00 |
| | Yin et al. | 49.22 | 46.15 | 44.01 | 42.30 | 40.94 | 39.67 | 38.32 | 37.16 | 35.88 |
| | Ours | 45.35 | 43.68 | **44.86** | **45.06** | **41.65** | **40.35** | **39.2** | **39.27** | **37.71** |

TABLE V: The PSNR values of the marked images with quality factor is 70 and embedded with different lengths of data using our method and three state-of-the-art methods.

| Image | Scheme | Payload | | | | | | | | |
|---|---|---|---|---|---|---|---|---|---|---|
| | | 2000 | 4000 | 6000 | 8000 | 10000 | 12000 | 14000 | 16000 | 18000 |
| Baboon | He et al. | 41.53 | 38.78 | 36.98 | 35.11 | 34.4 | 33.82 | 32.73 | 32.77 | 32.29 |
| | Xiao et al. | **50.82** | **47.20** | **44.73** | **42.97** | **41.53** | **40.24** | **39.12** | 38.11 | 37.06 |
| | Yin et al. | 50.56 | 47.02 | 44.67 | 42.83 | 41.29 | 39.90 | 38.54 | 37.37 | 36.38 |
| | Ours | 44.22 | 38.78 | 36.98 | 35.57 | 40.58 | 38.81 | 35.81 | **38.78** | **37.24** |
| Lena | He et al. | 50.68 | 46.82 | 43.90 | 41.71 | 41.26 | 40.62 | 39.93 | 39.36 | 38.62 |
| | Xiao et al. | **53.45** | 50.02 | 47.66 | **45.89** | **44.27** | **42.87** | **41.66** | 40.39 | 39.05 |
| | Yin et al. | 53.24 | 49.78 | 47.35 | 45.40 | 43.95 | 42.62 | 41.37 | 39.99 | 38.62 |
| | Ours | 50.87 | **53.31** | **49.98** | 45.26 | 43.22 | 41.88 | 40.96 | **42.31** | **40.83** |

TABLE VI: The PSNR values of the marked images with quality factor is 90 and embedded with different lengths of data using our method and three state-of-the-art methods.

| Image | Scheme | Payload | | | | | | | | |
|---|---|---|---|---|---|---|---|---|---|---|
| | | 2000 | 5000 | 8000 | 11000 | 14000 | 17000 | 20000 | 23000 | 26000 |
| Baboon | He et al. | 50.08 | 44.75 | 41.62 | 39.37 | 37.84 | 37.28 | 36.37 | 35.97 | 35.45 |
| | Xiao et al. | **55.38** | **50.58** | **47.63** | **45.37** | **43.62** | **42.17** | **40.99** | **39.89** | 38.77 |
| | Yin et al.. | 54.40 | 49.47 | 46.79 | 44.83 | 43.13 | 41.89 | 40.68 | 39.69 | 38.77 |
| | Ours | 48.71 | 44.75 | 41.62 | 39.67 | 37.98 | 37.20 | 36.47 | 35.94 | **39.10** |
| Lena | He et al. | 54.64 | 49.25 | 47.93 | 48.49 | 46.08 | 47.37 | 46.19 | 44.20 | 43.05 |
| | Xiao et al. | 58.80 | 54.96 | **52.77** | **51.24** | **49.82** | **48.62** | **47.42** | 46.21 | 44.90 |
| | Yin et al. | **59.32** | **55.02** | 52.54 | 50.76 | 49.21 | 47.96 | 46.76 | 45.64 | 44.47 |
| | Ours | 54.64 | 51.48 | 49.12 | 49.16 | 47.74 | 48.14 | 47.26 | **47.20** | **45.51** |

B. Test in the database

To verify the proposed generality, 50 images are randomly selected from the image database UCID [24], the final results are averaged, the test results are shown in figure 9, figure 10 and figure 11. Since Du et al's [23] method is lossless data hiding, we do not compare PSNR with it.

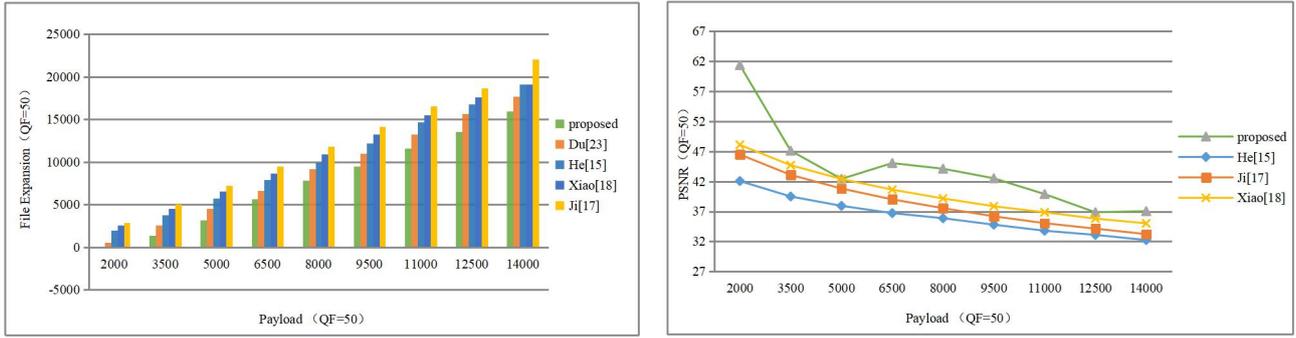

Fig.9    Average file size increments and PSNR of 50 images from the UCID image database(QF=50)

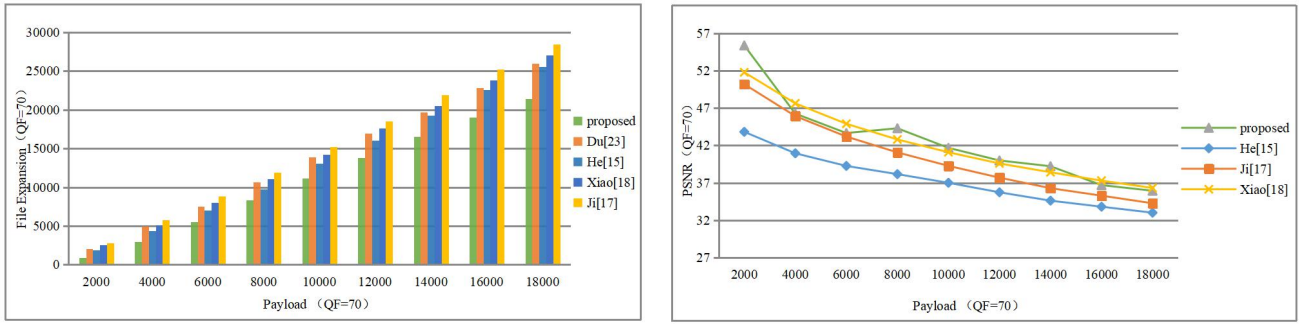

Fig.10    Average file size increments and PSNR of 50 images from the UCID image database(QF=70)

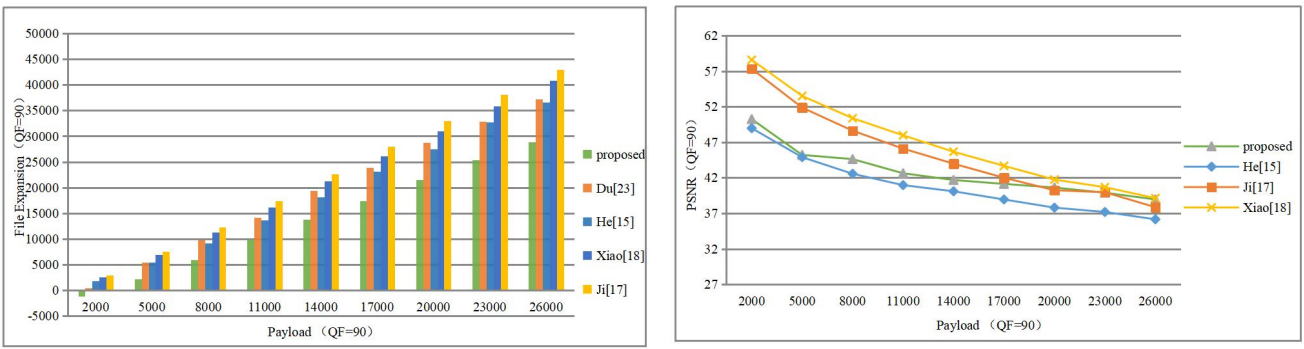

Fig.11    Average file size increments and PSNR of 50 images from the UCID image database(QF=90)

It can be seen from figure 9, when the quality factor are 50, 70 and 90, the file expansion of the proposed method is better than all methods obviously [15], [17], [18], [23]. It can be seen that He et al's [15] method achieves quite good result compared with Du et al's [23] method when quality factor are 70 and 90 in figure 10 and figure 11. The file expansion of proposed is negative when quality factor are 50 and 90 and payload is 2000 bits. The PSNR of the proposed method is always better than the other three methods [15], [17], [18] except the payload is 5000 when quality factor is 50, as can be seen from figure 9. The PSNR of the proposed method still has advantages when quality factor is 70 and payload are 2000, 8000, 10000, 12000, 14000, this advantage is not exist when quality factor is 90, but the PSNR of proposed method is always better than that of he et al's

[15] method, it can be seen from figure 10 and figure 11, the reason of this phenomenon is that the embedding cost of entropy coding domain is higher than that of DCT coefficient domain in most cases when the quality factor is 90, more secret bits are embedded in the DCT coefficient domain, resulting in visual quality distortion. We can see from the rate distortion curve in figure 9, figure 10 and figure 11 that the PNSR of Xiao et al's [18] method is the best among the three methods [15], [17], [18] based on DCT coefficient modification.

C. Effectiveness of the proposed payload distribution

Since the proposed method has the process of optimizing Huffman table, a simple experiment is designed to illustrate the effectiveness of the payload distribution algorithm. After all the secret data is embedded by He et al's method, the Huffman table is optimized to generate a new image. Experimental results show that the image generated by the proposed method is closer to the original image than the new image. The experiment results are shown in the Table VII, Table VIII, Table IX and Table X.

TABLE V: The file size changes in bits of the marked images and new images with quality factor is 50 and embedded with different lengths.

| Image | Scheme | Payload | | | | | | |
|---|---|---|---|---|---|---|---|---|
| | | 4000 | 5500 | 7000 | 8500 | 10000 | 11500 | 13000 |
| Baboon | He et al. | -1816 | -224 | **1160** | **2464** | **4424** | **6440** | **8000** |
| | Ours | -1816 | -224 | 1768 | 3200 | 4904 | 6842 | 8200 |
| Lena | He et al. | **1760** | 4064 | 5680 | 7656 | 9304 | 11312 | 13392 |
| | Ours | 1824 | **3160** | **5056** | **7376** | **9024** | **10664** | **12632** |

TABLE VI: The file size changes in bits of the marked images and new images with quality factor is 70 and embedded with different lengths.

| Image | Scheme | Payload | | | | | | |
|---|---|---|---|---|---|---|---|---|
| | | 4000 | 5500 | 7000 | 8500 | 10000 | 11500 | 13000 |
| Baboon | He et al. | 832 | 2376 | 4072 | 5672 | **8032** | 9800 | **12024** |
| | Ours | 832 | 2376 | **3984** | **5584** | 8776 | **9688** | 12312 |
| Lena | He et al. | **2936** | **4744** | 6656 | 8800 | 10624 | 12376 | 14280 |
| | Ours | 2936 | 5096 | **6672** | **8400** | **10352** | **11880** | **14024** |

TABLE VII: The PSNR values of the marked images and new images with quality factor is 50 and embedded with different lengths.

| Image | Scheme | Payload | | | | | | |
|---|---|---|---|---|---|---|---|---|
| | | 4000 | 5500 | 7000 | 8500 | 10000 | 11500 | 13000 |
| Baboon | He et al. | 37.74 | 35.84 | 34.79 | 33.98 | 33.53 | 32.8 | 32.35 |
| | Ours | 37.74 | 35.84 | **42.84** | **38.19** | **36.92** | **35.2** | **34.63** |
| Lena | He et al. | 41.80 | 40.01 | 38.97 | 38.34 | 37.70 | 36.84 | 36.29 |
| | Ours | **44.68** | **43.17** | **43.22** | **40.85** | **39.44** | **40.15** | **38.62** |

TABLE VIII: The PSNR values of the marked images and new images with quality factor is 70 and embedded with different lengths.

| Image | Scheme | Payload | | | | | | |
|---|---|---|---|---|---|---|---|---|
| | | 4000 | 5500 | 7000 | 8500 | 10000 | 11500 | 13000 |
| Baboon | He et al. | 39.01 | 35.99 | 34.91 | 34.49 | 34.23 | 33.79 | 32.97 |
| | Ours | **39.02** | **37.04** | **36.22** | **34.78** | **40.03** | **37.69** | **36.52** |
| Lena | He et al. | 45.26 | 44.15 | 41.99 | 41.36 | 40.52 | 40.22 | 39.68 |
| | Ours | **45.26** | **47.32** | **45.78** | **44.55** | **42.20** | **41.37** | **40.81** |

As can be seen from Table VII and Table VIII, the file expansion caused by the proposed method is close to that of the new image, and in the test of Lena image, the proposed method has more advantages especially. It can be seen from Table IX and Table X that the PSNR caused by the proposed method is still better than that of the new image, which is inevitable because the process of optimizing Huffman table does not cause any image visual quality distortion. Therefore, the test results in Table IX and Table X are the same as those in Table III and Table IV. It shows that the proposed payload distribution algorithm is effective.

## V. CONCLUSION

This paper proposes a multi-domain reversible data hiding in JPEG Images algorithm. We combine the method based on DCT coefficient domain modification with the method based on entropy coding domain modification, and propose the embedding cost, and according to the embedding cost, the payload is distributed to two domains, it is effective to reduce the file expansion and improve peak signal-to-noise ratio. Compared with the other four methods [15], [17], [18], [23], the proposed method has lower file expansion. The PSNR of the proposed method also has some advantages when the quality factor is low.

**REFERENCES**

[1]Ni Z, Shi Y Q, Ansari N, et al. Reversible data hiding[J]. IEEE Transactions on circuits and systems for video technology, 2006, 16(3): 354-362.

[2]J. Fridrich, M. Goljan, and R. Du, "Lossless data embedding—new paradigm in digital watermarking," EURASIP Journal on Advances in Signal Processing, vol. 2002, no. 2, p. 986842, 2002.

[3]M. U. Celik, G. Sharma, A. M. Tekalp, and E. Saber, "Lossless generalized-lsb data embedding," IEEE transactions on image processing, vol. 14, no. 2, pp. 253–266, 2005.

[4]J. Tian, "Reversible data embedding using a difference expansion," IEEE transactions on circuits and systems for video technology, vol. 13, no. 8, pp. 890–896, 2003.

[5]I.-C. Dragoi and D. Coltuc, "Local-prediction-based difference expansion reversible watermarking," IEEE Transactions on image processing, vol. 23, no. 4, pp. 1779–1790, 2014.

[6]X. Zhang, "Reversible data hiding with optimal value transfer," IEEE Transactions on Multimedia, vol. 15, no. 2, pp. 316–325, 2012.

[7]W. Zhang, X. Hu, X. Li, and N. Yu, "Recursive histogram modification: establishing equivalency between reversible data hiding and lossless data compression," IEEE transactions on image processing, vol. 22, no. 7, pp. 2775–2785, 2013.

[8]Y. Jia, Z. Yin, X. Zhang, and Y. Luo, "Reversible data hiding based on reducing invalid shifting of pixels in histogram shifting," Signal Processing, vol. 163, pp. 238–246, 2019.

[9]Fridrich J , Goljan M , Du R . Lossless data embedding for all image formats[J]. Spie Security & Watermarking of Multimedia Contents IV, 2002, 4675:572-583.

[10]Wang K, Lu Z M, Hu Y J. A high capacity lossless data hiding scheme for JPEG images[J]. Journal of systems and software, 2013, 86(7): 1965-1975.

[11]Xuan G , Shi Y , Ni Z , et al. Reversible data hiding for JPEG images based on histogram pairs[J]. 2007.

[12]Huang F , Qu X , Kim H J , et al. Reversible Data Hiding in JPEG Images[J]. IEEE Transactions

on Circuits & Systems for Video Technology, 2016, 26(9):1610-1621.

[13]Wedaj F T, Kim S, Kim H J, et al. Improved reversible data hiding in JPEG images based on new coefficient selection strategy[J]. EURASIP Journal on Image and Video Processing, 2017, 2017(1): 63.

[14]Hou D, Wang H, Zhang W, et al. Reversible data hiding in JPEG image based on DCT frequency and block selection[J]. Signal Processing, 2018, 148: 41-47.

[15]He J, Chen J, Tang S. Reversible Data Hiding in JPEG Images Based on Negative Influence Models[J]. IEEE Transactions on Information Forensics and Security, 2019.

[16]He J, Pan X, Wu H, et al. Improved block ordering and frequency selection for reversible data hiding in JPEG images[J]. Signal Processing, 2020, 175: 107647.

[17]Yin Z , Ji Y , Luo B . Reversible Data Hiding in JPEG Images with Multi-objective Optimization[J]. IEEE Transactions on Circuits and Systems for Video Technology, 2020, PP(99):1-1.

[18]Xiao M, Li X, Ma B, et al. Efficient reversible data hiding for JPEG images with multiple histograms modification[J]. IEEE Transactions on Circuits and Systems for Video Technology, 2020.

[19]Mobasseri B G, Berger R J, Marcinak M P, et al. Data embedding in JPEG bitstream by code mapping[J]. IEEE Transactions on image processing, 2009, 19(4): 958-966.

[20]Qian Z, Zhang X. Lossless data hiding in JPEG bitstream[J]. Journal of Systems and Software, 2012, 85(2): 309-313.

[21]Y. Qiu, H. He, Z. Qian, S. Li, and X. Zhang, "Lossless data hiding in JPEG bitstream using alternative embedding," Journal of Visual Communication and Image Representation, vol. 52, pp. 86–91, 2018.

[22]Du Y, Yin Z, Zhang X. Improved lossless data hiding for JPEG images based on histogram modification[J]. Computers, Materials & Continua, 2018, 55(3): 495-507.

[23]Du Y, Yin Z, Zhang X. High Capacity Lossless Data Hiding in JPEG Bitstream Based on General VLC Mapping[J]. IEEE Transactions on Dependable and Secure Computing, 2020.

[24]G. Schaefer and M. Stich, "Ucid: An uncompressed color image database," in Storage and Retrieval Methods and Applications for Multimedia 2004, vol. 5307. International Society for Optics

and Photonics, 2003, pp. 472–481.